\documentclass{aa}  

\usepackage[normalem]{ulem}
\usepackage{textcomp}
\usepackage{dcolumn}
\newcolumntype{d}[1]{D{.}{.}{#1}}  
\usepackage{graphicx}
\usepackage{xcolor,colortbl}
\usepackage{txfonts}
\usepackage{natbib}
\usepackage[colorlinks=true,urlcolor=blue,citecolor=blue,linkcolor=blue,breaklinks=true]{hyperref}
\begin{document}

\title{Solar active region evolution and imminent flaring activity through a color-coded visualization of photospheric vector magnetograms}

   \author{I. Kontogiannis
          \inst{1}
          \and 
          A.G.M. Pietrow\inst{1,2}
          \and
          M.K. Druett\inst{2}
          \and 
          E. Dineva\inst{2}
          \and 
          M. Verma\inst{1}
          \and
          C. Denker\inst{1}
          }

   \institute{Leibniz-Institut f\"ur Astrophysik Potsdam (AIP), Germany, An der Sternwarte 16 14482 Potsdam Germany
   \and
   Centre for Mathematical Plasma Astrophysics, Department of Mathematics, KU Leuven, Celestijnenlaan 200B, B-3001 Leuven, Belgium\\
              \email{ikontogiannis@aip.de}
             }

   \date{\today}

 
  \abstract
   {The emergence of magnetic flux, its transition to complex configurations, and the pre-eruptive state of active regions are probed using photospheric magnetograms.}
   {Our aim is to pinpoint different evolutionary stages in emerging active regions, explore their differences and produce parameters that could advance flare prediction, using color-coded maps of the photospheric magnetic field.}
   {The three components of the photospheric magnetic field vector are combined to create color-coded magnetograms (COCOMAGs). From these, the areas occupied by different color hues are extracted, creating appropriate time series (color curves). These COCOMAGs and color curves are used as proxies of the active region evolution and its complexity.}
   {The COCOMAGs exhibit a morphology relevant to typical features of active regions, such as sunspots, plages, and sheared polarity inversion lines. The color curves represent the area occupied by photospheric magnetic field of different orientation and contain information relevant to the evolutionary stages of active regions. During emergence, most of the region area is dominated by horizontal or highly inclined magnetic field, which gradually gives its place to more vertical magnetic field. In complex regions, large parts are covered by highly inclined magnetic fields, appearing as abrupt color changes in COCOMAGs. The decay of a region is signified by a domination of vertical magnetic field, indicating a gradual relaxation of the magnetic field configuration. The color curves exhibit varying degree of correlation with active region complexity. Particularly the red and magenta color curves, which represent strong, purely horizontal magnetic field, are good indicators of future flaring activity.}
   {Color-coded magnetograms facilitate a comprehensive view of the evolution of active regions and their complexity. They offer a framework for the treatment of complex observations and can be used in pattern recognition, feature extraction and flare prediction schemes.}


\titlerunning{Color-coded magnetograms}
\authorrunning{Kontogiannis et al.}
\maketitle
%

\section{Introduction}\label{intro}

The emergence of magnetic flux in the solar atmosphere often leads to the formation of highly complex magnetic field configurations, called active regions \citep{2015LRSP...12....1V}. The interaction of the emerging magnetic field with the convection and the flow field in the photosphere results in numerous manifestations in the continuum, and thus active regions often contain groups of sunspots, pores, and plages \citep[see definitions in][]{Cretignier23}, with varying complexity and size \citep{1919ApJ....49..153H}. Active regions host highly energetic phenomena such as Ellerman bombs \citep{Ellerman1917,Georgoulis02,Watanabe11}, flares \citep{Hirayama74,Fletcher11}, and coronal mass ejections \citep[CMEs; see e.g.,][for comprehensive reviews of their properties]{webb12}. Both flares and CMEs are facets of the solar eruptive activity and cause adverse space weather effects \citep{2017EastwoodEconomics}. Therefore, providing alternative ways to describe the evolution of active regions has the potential to reveal the details of flux emergence and provide potentially useful tools for the prediction of eruptions.

Emerging magnetic flux initially appears as patches of opposite polarities (seen as circular polarization with opposite signs), which start to grow and separate, while in between linear polarization betrays the presence of the horizontal magnetic field that connects the two polarities. As the polarities separate, the region in between fills with the so-called serpentine field structure with alternating opposite polarities separated with horizontal magnetic field \citep{2004Pariat}. This undulatory structure has also been revealed in the highest resolution possible to date \citep{2023CampbellSerpentine} and seen in small-scale flux emergence events \citep{2020Kontogiannis}. If the emerging flux develops further into complex regions, shearing and twisting motions become more evident through features such as magnetic tongues \citep{2003A&A...397..305L} and complex magnetic polarity inversion lines (PILs), which demarcate regions of strong interaction with shearing flows and flux cancellation \citep{toriumi17,2018Verma,chintzoglou19,Pietrow23}. In fact, PILs are the most prominent indicators of imminent flare-productivity and eruptive activity \citep{schrijver07}, implying a complex magnetic morphology. These manifestations rely highly on the horizontal component of the photospheric magnetic field, therefore a simultaneous monitoring of all three components of the magnetic field could be important for studies focusing on active region evolution.

\begin{table*}
\caption{The sample of 33 active regions used for the statistics presented in Figs~\ref{fig:corrs} and \ref{fig:flare_probs}. For each region the National Oceanic and Atmospheric Administration (NOAA) identifier, the start and end date and heliographic longitude of each SHARP data set and the resulting number of processed magnetograms (the number of points in the time series) are presented. The last column list the total number of flares associated with each region, based on the SWAN data set \citep{2020NatSD...7..227A}.}             
\label{table:2}  
\centering          
\begin{tabular}{l l l l l l r  l }     
\hline\hline       
No. & NOAA & Start date [UT] & End date [UT] & Start HL [\textdegree] & End HL [\textdegree] & Frames & Flares \\
\hline
     \hphantom{0}1 &  11072  &  20 \makebox[0.7cm]{May} 2010 16:22  &  29 \makebox[0.7cm]{May} 2010 12:34  &  --35.76  &  \hphantom{--}90.09  & 1050 & \hphantom{0}4\,B, \hphantom{ 000 C, 00 M, 0 X}\\
      \hphantom{0}2 &  11079  &  08 \makebox[0.7cm]{Jun} 2010 03:34  &  13 \makebox[0.7cm]{Jun} 2010 13:46  &  \hphantom{--}12.10  &  \hphantom{--}89.36  & 589 & \hphantom{0}1\,B, \hphantom{000 C,} \hphantom{0}1\,M \hphantom{, 0 X}\\
      \hphantom{0}3 &  11117  &  18 \makebox[0.7cm]{Oct} 2010 19:34  &  01 \makebox[0.7cm]{Nov} 2010 15:58  &  --88.29  &  \hphantom{--}91.87  & 1502 & 55\,B, \hphantom{00}5\,C \hphantom{, 00 M, 0 X}\\
      \hphantom{0}4 &  11158  &  10 \makebox[0.7cm]{Feb} 2011 21:58  &  21 \makebox[0.7cm]{Feb} 2011 07:34  &  --40.26  &  \hphantom{--}92.56  & 1247 & \hphantom{0}1\,B, \hphantom{0}56\,C, \hphantom{0}5\,M, 1\,X\\
      \hphantom{0}5 &  11166  &  02 \makebox[0.7cm]{Mar} 2011 08:34  &  10 \makebox[0.7cm]{Mar} 2011 23:46  &  --84.33  &  \hphantom{--}33.19  & 852 & \hphantom{0}4\,B, \hphantom{0}27\,C, \hphantom{0}4\,M, 1\,X \\
      \hphantom{0}6 &  11267  &  04 \makebox[0.7cm]{Aug} 2011 10:10  &  12 \makebox[0.7cm]{Aug} 2011 11:22  &  --46.62  &  \hphantom{--}58.76  & 963 & \hphantom{0}1\,B, \hphantom{00}4\,C \hphantom{, 00 M, 0 X}\\
      \hphantom{0}7 &  11273  &  16 \makebox[0.7cm]{Aug} 2011 13:10  &  21 \makebox[0.7cm]{Aug} 2011 16:34  &  --18.69  &  \hphantom{--}52.27  & 618 & \\
      \hphantom{0}8 &  11283  &  03 \makebox[0.7cm]{Sep} 2011 23:58 & 09 \makebox[0.7cm]{Sep} 2011 23:46  &  --14.15  &  \hphantom{--}66.03  & 719  & \hphantom{0}7\,B, \hphantom{0}13\,C, \hphantom{0}5\,M, 2\,X\\
      \hphantom{0}9 &  11327  &  18 \makebox[0.7cm]{Oct} 2011 23:58  &  28 \makebox[0.7cm]{Oct} 2011 16:34  &  --38.71  &  \hphantom{--}87.77  & 1107 & \\
      10 &  11429  &  04 \makebox[0.7cm]{Mar} 2012 23:58 &  08 \makebox[0.7cm]{Mar} 2012 23:46 &  --33.82  &  \hphantom{--}19.44  & 467 &  \hphantom{0}2\,B, \hphantom{0}32\,C, 14\,M, 2\,X\\
      11 &  11431  &  04 \makebox[0.7cm]{Mar} 2012 11:10  &  10 \makebox[0.7cm]{Mar} 2012 09:22  &  \hphantom{--}14.62  &  \hphantom{--}94.20  & 681 & \\
      12 &  11476  &  04 \makebox[0.7cm]{May} 2012 12:58  &  17 \makebox[0.7cm]{May} 2012 22:34  &  --84.68  &  \hphantom{--}89.44  & 1477 & 13\,B, \hphantom{0}89\,C, 11\,M \hphantom{, 0 X}\\
      13 &  11515  &  26 \makebox[0.7cm]{Jun} 2012 03:58  &  10 \makebox[0.7cm]{Jul} 2012 03:10  &  --81.46  &  \hphantom{--}88.61  & 1668 & \hphantom{0}2\,B, \hphantom{0}73\,C, 30\,M, 1\,X \\
      14 &  11726  &  19 \makebox[0.7cm]{Apr} 2013 04:22  &  27 \makebox[0.7cm]{Apr} 2013 09:10  &  --19.41  &  \hphantom{--}88.68  & 924 & 11\,B, \hphantom{0}56\,C, \hphantom{0}1\,M \hphantom{, 0 X}\\
      15 &  11884  &  26 \makebox[0.7cm]{Oct} 2013 06:58  &  08 \makebox[0.7cm]{Nov} 2013 10:46  &  --83.82  &  \hphantom{--}88.84  & 1571 & \hphantom{0}1\,B, \hphantom{0}15\,C, \hphantom{0}4\,M \hphantom{, 0 x}\\
      16 &  11890  &  02 \makebox[0.7cm]{Nov} 2013 23:58  &  11 \makebox[0.7cm]{Nov} 2013 23:46  &  --68.05  &  \hphantom{--}53.28  & 1072 & \hphantom{00\,B, 0}46\,C, \hphantom{--}5\,M, 3\,X \\
      17 &  12003  &  09 \makebox[0.7cm]{Mar} 2014 15:34  &  16 \makebox[0.7cm]{Mar} 2014 22:10  &  --07.31  &  \hphantom{--}89.50  & 765 & \hphantom{00 B, 00}9\,C \hphantom{00 M, 0 x}\\
      18 &  12089  &  10 \makebox[0.7cm]{Jun} 2014 20:22  &  19 \makebox[0.7cm]{Jun} 2014 22:10  &  --31.16  &  \hphantom{--}90.45  & 1082 &  \hphantom{0}1\,B, \hphantom{0}11\,C, \hphantom{0}1\,M \hphantom{, 0 x}\\
      19 &  12118  &  17 \makebox[0.7cm]{Jul} 2014 16:34  &  21 \makebox[0.7cm]{Jul} 2014 04:10  &  \hphantom{--}13.30  &  \hphantom{--}59.95  & 419 & \\
      20 &  12119  &  18 \makebox[0.7cm]{Jul} 2014 09:34  &  26 \makebox[0.7cm]{Jul} 2014 04:46  &  --23.16  &  \hphantom{--}84.18  & 867 & \hphantom{0}1\,B \hphantom{,000 c, 00 m, 0 x}\\
      21 &  12192  &  16 \makebox[0.7cm]{Oct} 2016 09:10  &  30 \makebox[0.7cm]{Oct} 2016 18:22 &  --82.71  &  \hphantom{--}88.92  & 924 &  \hphantom{00 b, 0}74\,C, 32\,M, 6\,X \\ 
      22 &  12205  &  04 \makebox[0.7cm]{Nov} 2014 23:58  &  08 \makebox[0.7cm]{Nov} 2014 23:46  &  --63.56  &  --10.20  & 480 &  \hphantom{00 b, 0}42\,C, 13\,M, 1\,X\\
      23 &  12219  &  24 \makebox[0.7cm]{Nov} 2014 06:10  &  02 \makebox[0.7cm]{Dec} 2014 20:58  &  --31.45  &  \hphantom{--}89.82  & 1003 &  \hphantom{00 b, 00}7\,C \hphantom{, 00 m, 0 x}\\
      24 &  12234  &  09 \makebox[0.7cm]{Dec} 2014 06:34  &  18 \makebox[0.7cm]{Dec} 2014 06:46  &  --42.95  &  \hphantom{--}89.77  & 1072 &  \hphantom{00 b, 00}4\,C \hphantom{, 00 m, 0 x}\\
      25 &  12242  &  17 \makebox[0.7cm]{Dec} 2014 23:58  &  21 \makebox[0.7cm]{Dec} 2014 23:46 &  \hphantom{--}55.65  &  \hphantom{--}87.65  & 475 &  \hphantom{00 b, 0}55\,C, \hphantom{0}6\,M, 1\,X \\
      26 &  12257  &  04 \makebox[0.7cm]{Jan} 2015 03:46  &  14 \makebox[0.7cm]{Jan} 2015 14:46  &  --50.51  &  \hphantom{--}89.60  & 1252 & \hphantom{00 b, 0}23\,C, \hphantom{0}3\,M \hphantom{0 x}\\
      27 &  12271  &  24 \makebox[0.7cm]{Jan} 2015 06:22  &  02 \makebox[0.7cm]{Feb} 2015 14:22  &  --32.58  &  \hphantom{--}87.96  & 1087 & \hphantom{0}1\,B, \hphantom{00}4\,C \hphantom{, 00 m, 0 x}\\
      28 &  12297  &  08 \makebox[0.7cm]{Mar} 2015 23:58  &  12 \makebox[0.7cm]{Mar} 2015 23:46 &  --45.99  &  \hphantom{--}07.01  & 419 &  17\,B, 100\,C, 23\,M, 1\,X \\
      29 &  12339  &  04 \makebox[0.7cm]{May} 2015 16:34  &  18 \makebox[0.7cm]{May} 2015 17:58  &  --84.50  &  \hphantom{--}89.56  & 1636 & \hphantom{0}3\,B, \hphantom{0}56\,C, \hphantom{0}3\,M, 1\,X\\
      30 &  12371  &  15 \makebox[0.7cm]{Jun} 2015 14:34  &  28 \makebox[0.7cm]{Jun} 2015 23:58  &  --86.64  &  \hphantom{--}90.45  & 1602 & 13\,B, \hphantom{0}40\,C, \hphantom{0}7\,M \hphantom{, 0 x}\\
      31 &  12422  &  22 \makebox[0.7cm]{Sep} 2015 07:58  &  03 \makebox[0.7cm]{Oct} 2015 22:46  &  --59.55  &  \hphantom{--}87.49  & 924  & 12\,B, \hphantom{0}62\,C, 18\,M \hphantom{, 0 x} \\
      32 &  12543  &  07 \makebox[0.7cm]{May} 2016 16:58  &  16 \makebox[0.7cm]{May} 2016 17:58  &  --30.72  &  \hphantom{--}90.04  & 1043 & \hphantom{0}7\,B, \hphantom{00}5\,C \hphantom{, 00 m, 0x}\\
      33 &  12673  &  28 \makebox[0.7cm]{Aug} 2017 08:58  &  10 \makebox[0.7cm]{Sep} 2017 11:10  &  --81.77  &  \hphantom{--}89.12  & 1431 &  \hphantom{0}1\,B, \hphantom{0}54\,C, 26\,M, 4\,X\\
\hline                  
\end{tabular}
\end{table*}

Ever since photospheric magnetograms became easily accessible, this morphological complexity of active regions has been parameterised using appropriate metrics derived from the magnetic field measurements \citep[see][for a detailed account of these efforts and some of the future challenges]{kontogiannis23}. In summary, these metrics can be physical quantities (magnetic flux, electric current, magnetic energy, and various types of helicity) or intuitive quantities which encapsulate the accumulated empirical and theoretical knowledge on eruption initiation. This approach is not limited only to magnetograms, but it can also be extended to coronal observations \citep[see e.g.,][]{2020arXiv201106433G,2024arXiv240704567H}. Manipulating the information contained in observations in creative and intuitive ways is crucial to finding concise methods of parameterizing the evolution of active regions towards eruptions and, thereby, improve forecasts \citep[see, e.g.,][]{2021JSWSC..11...39G}. Moreover, different approaches can also bring out hitherto overlooked aspects of the magnetic field evolution. Recognition of relevant patterns could then be done either by experts, since humans have a highly developed visual perception and pattern recognition \citep{1101675601,c2008digital}, or by properly trained machine learning (ML) algorithms.

A recent example of bringing out important information is the work by \citet{Denker19}, who introduced the Background Subtracted Activity Maps (BaSAMs) to highlight regions of intense variability within time series of images. Application to high resolution images of pores at different atmospheric layers revealed regions of prominent changes in active regions and facilitated establishing the connectivity of intense variability between various heights \citep{2023Kamlah}. This method can also be used to pinpoint significant variations in time series of spectral imaging, thus offering a way to parse data summarily and reduce dimensionality \citep{2023RNAAS...7..224D}. Complex spectral data can also be summarised and analysed through grouping with other ``similar'' profiles, as judged through some heuristic. One example of this method is the K-means clustering method \citep{1967MacqueenKmeans}, which has seen extensive use in solar physics, as was concisely summarised in a recent paper by \citet{2023MoeK-means}. Another is the t-distributed stochastic neighbor embedding method \citep[t-SNE,]{2008VanDerMaatenTSNE} recently employed to solar physics applications by \citet{2021VermaTSNE}.

An alternative way to examine spectral imaging data was proposed by \citet{2022RASTI...1...29D}. The COlor COllapsed PLOTting software (COCOPLOTS) is projecting a spectrum into the RGB vector, thus transforming spectral information into color. Therefore, color-coded maps resulting from spectral imaging data sets can then be used to study dynamic evolution and pinpoint features of interest. Both COCOPLOTS and BaSAMS were used to locate the observed chromospheric changes as a result of two X-class flares \citep{Pietrow23}. Subsequent analysis of the localized regions of interest detected through COCOPLOT facilitated the categorization of the diverse structures found in flare ribbons \citep{2024A&A...685A.137P}. A similar color-coding process was employed by \citet{Osborne22} to study the wavelength-dependent changes induced by flaring to contribution functions. 

Motivated by these works, in this study we propose an alternative method to visualize photospheric vector magnetograms, aiming to showcase new ways of studying the evolution of active regions and assessing their eruptive potential. The proposed method combines the three cartesian components of the photospheric magnetic field using the methodology of COCOPLOTS. The resulting COlor-COmbined MAGnetograms (COCOMAGs) encapsulate information from all three components, enabling their simultaneous monitoring during flux emergence and decay. The information in the RGB color space is used to create time series, which parameterise the evolution of the active regions (or any type of spatial distribution of the magnetic field vector) in terms of different colors. We then use the proposed methodology to explore aspects of flux emergence and magnetic complexity development as well as examine the potential use of COCOMAGs in the prediction of imminent flaring activity. 


\section{Data and Analysis}

COCOPLOTS \citep[COCOPLOTS,][]{2022RASTI...1...29D} offers a quick way to summarize spectral information by making use of the RGB color space, or a similar colorblind friendly map \citep{Osborne22}. Typically, this is applied to spectral lines, which are convolved with three equally spaced Gaussian wavelength filters, creating three filtergrams centered on the line center and the red and blue line wings. When combined into an RGB image, the relative pixel value of each of these three filtergrams is expressed as a color, which acts as a summary of the profile in the spectral dimension. For example, a purple pixel represents a typical absorption profile with higher intensities in the wings (red and blue filters) than in the line core (green filter) and a non-shifted emission profile will look green \citep[See section 2.2 of][for further examples]{2022RASTI...1...29D}. 

The concept of color-collapsed plots can be generalized to include any type of information or vector of at least three elements, convolved with three filters. In this case, we apply the method to the Bx, By, and Bz components of the photospheric magnetic field. We use the vector magnetograms provided by the Helioseismic and Magnetic Imager \citep[HMI;][]{hmischerrer,hmischou} onboard the Solar Dynamics Observatory \citep[SDO][]{sdo}. For the purposes of our study, we use the Space Weather HMI Active Region Patches \citep[SHARP;][]{bobra14}, which contain cut-outs around high polarization regions such as active regions and plages. More specifically we utilize the definitive Cylindrical Equal Area (CEA) product, which contains the three components of the magnetic field, remapped and de-projected to the solar disk center. The data are accompanied by an assembly of quantities suitable for operational space weather prediction and are appropriate for the derivation of additional parameters that characterize the complexity and eruptive potential of active regions \citep{kontogiannis23}. A sample of 33 randomly selected active regions of Solar Cycle 24 was gathered, which exhibited varying levels of flaring activity (Table~\ref{table:2}).

As there are no spectra to convolve, we use the COCOPLOT method to simply scale the absolute values of each magnetogram pixel to a value between 0 and 255 and then combine the three images into an RGB figure. The maximum value is set at two times the standard deviation of the map, with everything above it being clipped, and the minimum value is set to 0\,G. In addition, these COlor COmbined MAGnetograms (COCOMAGs) can be used to study the evolution of certain regions by means of so-called ``color spheres'', which are three-dimensional shapes in RGB space that can be used to isolate pixels with certain colors from the color plots. Overall, the approach enables also feature extraction by means of color manipulation in RGB or via transformation to other color spaces such as Hue-Saturation-Value (HSV). 

Examples of how different colors are associated with diverse magnetic information are shown in Table~\ref{table:1}. The color coordinates of the fully saturated primary (red, green, blue) and secondary (cyan, magenta, yellow) colors of the RGB color system, along with white and black are given in terms of scaled values of the magnetic field vector components. The three primary colors correspond to pixels where one of the components is maximum (according to the conditions described earlier) and the rest are zero. For each of the secondary colors, which are produced by combinations of the primary ones, two of the components are maximum and the third equal to zero. Thus, magenta contains contributions only from the horizontal components, while cyan and yellow are combinations of the vertical component and one of the two horizontal ones. 

Using the absolute values of the magnetic field allows us to exploit the full 8-bit range to map the strength of magnetic field components. The implication is that the polarity information is missing from the maps but they still retain significant geometric information of the magnetic field vector and present it concisely and intuitively. This will be evident in the following sections by the fact that COCOMAGs bring out important structures and morphological features of active regions. 

For the implementation of the method we did not choose a fixed and invariable magnetic field value range to avoid unnaturally dark or saturated maps at the end/beginning of the time series. The variable scaling of the magnetic field through these adaptive thresholds allows to infer properties of the magnetic field vector at a glance, with no other prior assumptions. Setting the minimum value to 0\,G ensures that the primary and secondary saturated colors correspond truly to pixels where respectively two or one of the magnetic field components are zero. The choice of $2\sigma$ results in a cut-off ranging from 20 to 800\,G, depending on the emergence stage and the size of the active region. The selected cut-off ensures that that the active region is always enclosed as it grows and that the upper range of the RGB coordinates always correspond to the strongest magnetic field. At the early stages of emergence, the low cut-off value corresponds mostly to the stochastic noise of the three components, the combination of which results mostly to low saturation colors. This choice does not introduce high values in the beginning of the derived color curves (see the description of color maps and curves in the following section). Although the results presented in the following sections were produced using the aforementioned compression of the magnetic field values, the method can be adapted to facilitate different needs and types of data.    

\begin{table}
\caption{Color coordinates and magnetic field components in the COCOMAG representation}             
\label{table:1}      
\centering          
\begin{tabular}{c c c c }     
\hline\hline       
 Color & RGB coordinates & Strong components \\

\hline     
White & [255, 255, 255] & Bx, Bz, By \\  

 \cellcolor{red}\textcolor{white}{Red} & [255, \hphantom{00}0, \hphantom{00}0] & Bx \hphantom{00, 000} \\
 \cellcolor{green} Green & [\hphantom{00}0, 255, \hphantom{00}0] & Bz \\
 \cellcolor{blue}\textcolor{white}{Blue} & [\hphantom{00}0, \hphantom{00}0, 255] & \hphantom{00, 00, }By \\
 \cellcolor{cyan}Cyan & [\hphantom{00}0, 255, 255] & \hphantom{00, }Bz, By \\
 \cellcolor{magenta}\textcolor{white}{Magenta} & [255, \hphantom{00}0, 255] & Bx\hphantom{, 00, }By \\
 \cellcolor{yellow} Yellow & [255, 255, \hphantom{00}0] & Bx, Bz\hphantom{, 00} & \\
   \cellcolor{black}\textcolor{white}{Black} & [\hphantom{00}0, \hphantom{00}0, \hphantom{00}0] & \\
\hline                  
\end{tabular}
\end{table}

\begin{figure*}[htp!]
\centering
\includegraphics[width=\hsize]{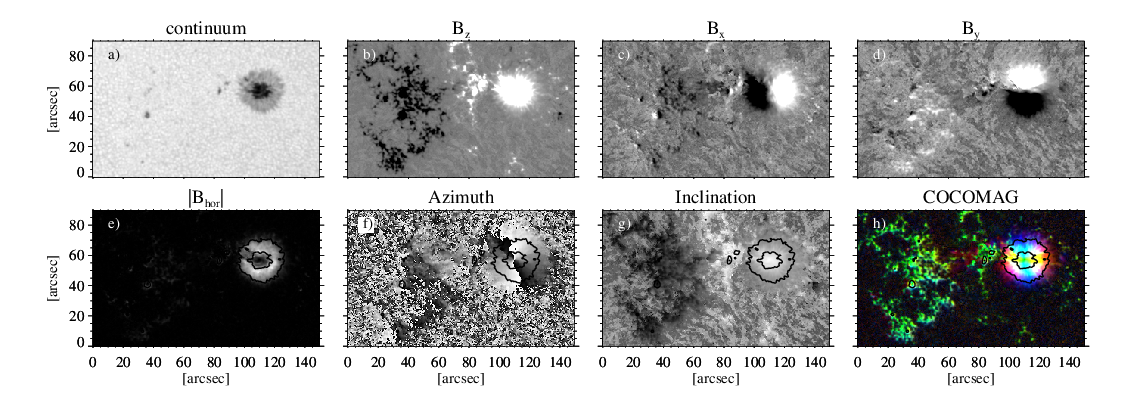}
\caption{Active region NOAA\,11072. The upper row shows information contained in the SHARPS CEA data, namely, (a) the maps of the continuum intensity, provided here for context, (b) the vertical component, $B_{z}$, and (c and d) the horizontal $B_{x}$ and $B_{y}$ components of the photospheric magnetic field. The magnetic field values were scaled to $\pm$500\,G for the vertical component and $\pm$400\,G for the horizontal ones. The lower row contains (e) the measure of the horizontal field $B_{hor}$, (f) the azimuth ranging from $-180\degr$ to $180\degr$, (g) the inclination ranging between $-90\degr$ and $90\degr$, and (h) the COCOMAG produced by $B_{x}$, $B_{y}$, and $B_{z}$. Contours of the continuum intensity were added to mark the boundaries of the umbra and the penumbra of the spot for context. }
\label{fig:noaa11072}
\end{figure*}

\section{Results}

\subsection{COCOMAGs and the magnetic field vector}

An example COCOMAG is shown in Fig.~\ref{fig:noaa11072}, comprising a snapshot from the evolution of the active region NOAA\,11072, several days after its emergence, containing a symmetric round spot (panel a). Visible in the $B_{z}$ map (panel b) is the strong, positive magnetic field of the umbra, which weakens towards the penumbra. Next to it smaller scattered polarities are found, mainly of negative magnetic polarity. The maps of $B_{x}$ and $B_{y}$ (panels c and d) indicate the horizontal direction of the sunspot magnetic field: the sign of $B_{x}$ ($B_{y}$) changes midway the sunspot, from positive to negative along the horizontal (vertical) direction and, as a result, the sunspot region appears, in each panel, divided in two sectors. The components of the magnetic field can be used to derive the total horizontal component, $B_\mathrm{hor} = \sqrt{B_{x}^2+B_{y}^2}$, the azimuth $\phi = \arctan(B_{y}/B_{x})$ and the inclination of the magnetic field $B_{z}/B_\mathrm{hor}$. It should be noted that the two latter are in fact the ones provided in the non-CEA version of SHARPS, since the azimuth and the inclination are the typical products of spectropolarimetric inversions. Fig.~\ref{fig:noaa11072}e shows that the stronger horizontal magnetic field is found at the penumbra, while it is predominantly weak over the small-scale magnetic fields of the plage. The azimuth map shows the 180\degr\ discontinuity that traverses the sunspot vertically, whereas the inclination map shows the transition from positive to negative inclinations (white and black, correspondingly) over the two well-separated magnetic polarities of the active region.

The COCOMAG of the region (Fig.~\ref{fig:noaa11072}h) provides in one colored map the aforementioned information (except for the polarity of the components). The white parts of the umbra and inner penumbra region correspond to regions where all three components, $B_{x}$, $B_{y}$, $B_{z}$ are higher than the $2\sigma$ cut-off. The peripehery of the penumbra as well as the surrounding region, where the vertical component is weaker, transitions from red to magenta, and blue, depending on which of the two horizontal components is prevalent. For the same reason, the sunspot is crossed by two roughly perpendicular stripes, a vertical yellow and a horizontal cyan one, reflecting the division of the sunspot seen in the $B_{x}$ and $B_{y}$ maps. Since yellow is produced by combining green and red (i.e., $B_{z}$ and $B_{x}$, see Table~\ref{table:1}), this strip runs along the sign-reversal line of $B_{y}$ (panel c in Fig.~\ref{fig:noaa11072}), where this component is zero or very weak. Similarly, the cyan strip in panel h, Fig.~\ref{fig:noaa11072} marks the region where the $B_{y}$ component is close to zero and corresponds to the sign-reversal region of $B_{x}$ seen in panel d, Fig.~\ref{fig:noaa11072}, and corresponds to the azimuth discontinuity (panel f in Fig.~\ref{fig:noaa11072}). Many of the small-scale magnetic elements of the active region are seen in green (predominantly vertical photospheric magnetic field) with parts seen in blue/cyan. Due to the higher noise level of the horizontal components, the noise pattern in the COCOMAG appears blue, red, and magenta.

\begin{figure}[htp!]
\centering
\includegraphics[width=\hsize]{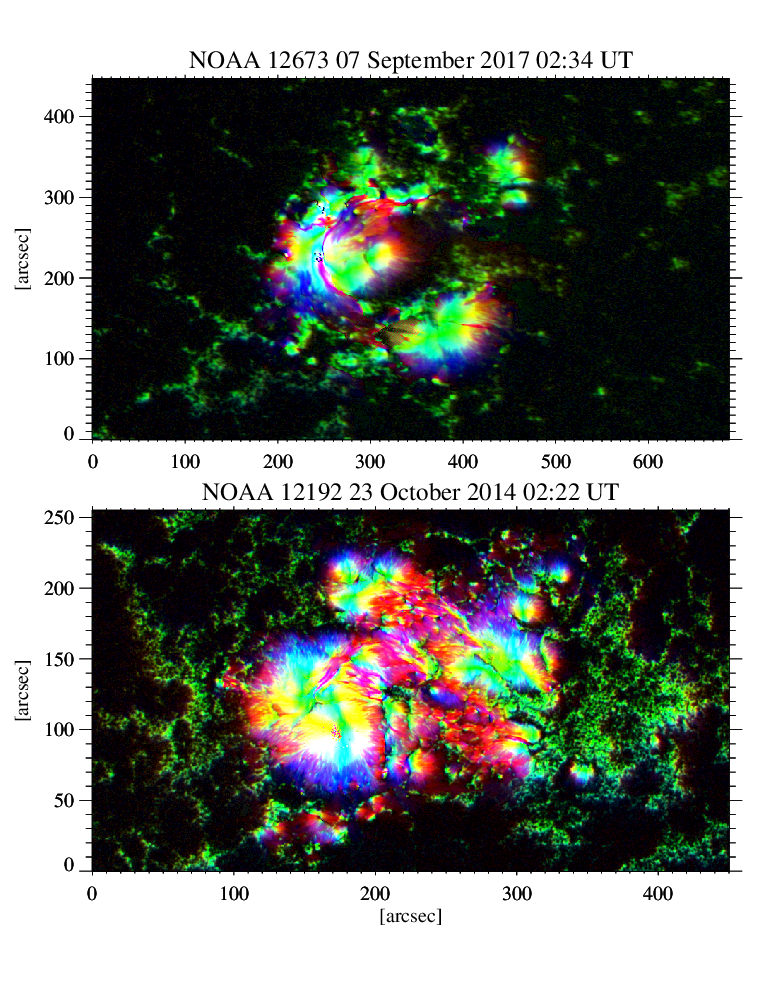}
\caption{COCOMAGs for active regions NOAA\,12673 (top) and NOAA\,12192 (bottom). }
\label{fig:examples}
\end{figure}

In the more general case of a highly complex region with asymmetric sunspots and overall magnetic field distribution, the appearance of sunspots and the magnetic field distribution differs. In Fig.~\ref{fig:examples}, we present two such examples, NOAA\,12673 and 12192, to illustrates how COCOMAGs can be used to probe the morphological complexity of active regions. Both of them were complex $\delta$-regions \citep{1965AN....288..177K} with high flare productivity. Due to their strong and highly deformed flux systems, their COCOMAGs are dominated by hues around the red, magenta, yellow and cyan colors, indicative of the strong presence of horizontal components in the photospheric magnetic field. In active region NOAA\,12673 (Fig.~\ref{fig:examples} top) a thin magenta strip crosses the magnetic field distribution, indicating the region of interaction where a PIL has formed, as a result of polarity collision \citep{2018Verma}. Consequently, the neighboring polarities have developed considerable twist and shear, hence the abrupt transition between different colors at these locations, depending on the local orientation of the magnetic field vector. Active region NOAA\,12192 (Fig.~\ref{fig:examples} bottom) is so far the region with the largest sunspot area since the beginning of Solar Cycle 24. Its COCOMAG exhibits similarities and differences with that of active region NOAA\,12673. The distorted procession of colors over the main polarities as well as the similarly distorted cyan strips crossing them indicate the existence of twist. The regions between the main footpoints also exhibit strong photospheric components and twist, evident as a smooth transition between different colors. However, unlike active region NOAA\,12673, there are no abrupt color changes, i.e., there is no clearly developed/detectable PIL. This result is in line with reports of relatively weaker non-potentiality of the region, which resulted to a low CME output, despite its strong flaring \citep{2015Sun}. 

\begin{figure*}[htp!]
\centering
\includegraphics[width=\hsize]{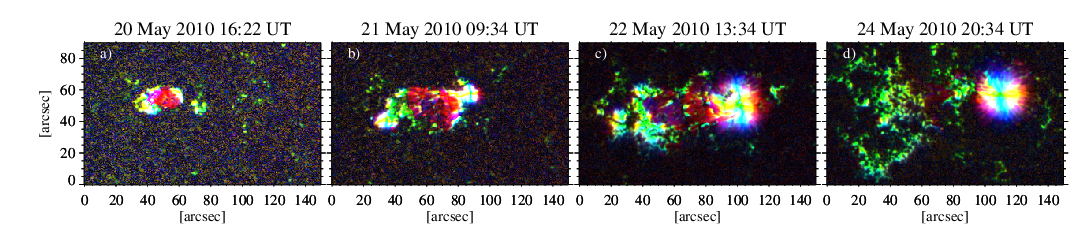}
\caption{Evolution of active region NOAA\,11072 in COCOMAG. }
\label{fig:evo11072}
\end{figure*}

\begin{figure*}[htp!]
\centering
\includegraphics[width=\hsize]{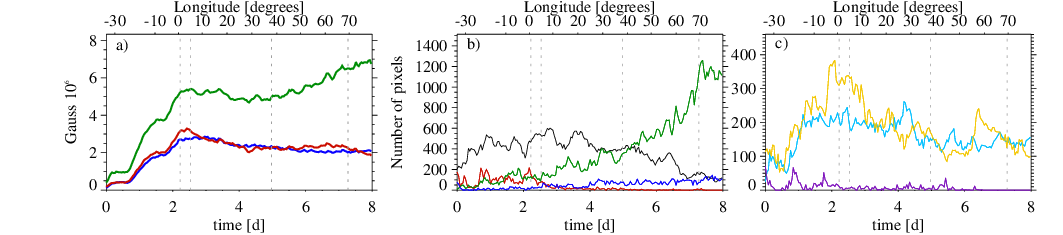}
\caption{A comparative overview of the evolution of the magnetic flux components and the color curves for active region NOAA\,11072. a) Total unsigned $B_{z}$ (green), $B_{x}$ (red), and $B_{y}$ (blue), b) total number of pixels with red, green, blue and white color (red, green, blue and black, respectively), and c) total number of pixels with cyan, magenta and yellow color. The vertical dashed lines indicate the times of B-class flares.}
\label{fig:lc11072}
\end{figure*}

The three exemplary cases of active regions presented so far showcase the capability of the COCOMAGs to reveal fundamental characteristics of the magnetic field vector. Even lacking polarity information, the morphology of the color-coded maps reflects intricate details regarding the complexity of active regions, ranging from symmetric sunspots with clear yellow and cyan strips, to those with clearly formed sunspots but with distorted strips indicating twist and, finally to those with absence of symmetry and abrupt changes of colors, with thin strips of magenta and red. 

\subsection{COCOMAGs and the emergence of magnetic flux}

Next, we study the dynamic evolution from emergence to decay in terms of COCOMAGs, starting with the familiar example of active region NOAA\,11072 (Fig.~\ref{fig:evo11072}). The COCOMAG in Fig.~\ref{fig:evo11072}a corresponds to the first instance of emergence provided by the SHARP data. The region starts as a compact colored patch, consisting of two separating crescent-shaped green-yellow-cyan patches both of them associated with strong vertical magnetic field, in combination with strong $B_{x}$ and to a lesser extent $B_{y}$. The red patch between the two main polarities indicates the emergence of horizontal magnetic field, predominantly in the $x$-direction. The dipole orientation aligns with the horizontal axis of the map, therefore, in all SHARP data red colors prevail over blue ones. The following day, on 21 May 2010, more of the horizontal part of the structure appears and the two polarities separate. They start to build up and contain more white, with the yellow-cyan strips in between indicating the formation of roughly circular pores/spots. There is still red between the two polarities, but some of the parts are also cyan, as the polarities also start to separate in the $y$-direction and significant $B_{y}$ component appears. On 22 May 2010, the positive polarity, located to the right, has developed the well-formed sunspot, already described in Fig.~\ref{fig:noaa11072}. The negative polarity to the left is more dispersed, with the rather vertical magnetic fields of the more compact parts of the polarity (green patches) surrounded by more inclined ones (cyan, yellow, and white in the periphery). There is still considerable emergence of magnetic flux betrayed by the red and magenta regions, which results in more flux accumulation in the two polarities of the region. Notable also are the mixed, small-scale, magnetic polarities in between the main polarities, since the magenta and red regions are separated by green patches at solar x ~ 70\arcsec. Two days later, the active region has the appearance seen in Fig.~\ref{fig:evo11072}d and described in Fig.~\ref{fig:noaa11072}, with a well-formed, well-separated, symmetrical sunspot at the right and a plage dominated by green pixels at the left.

In Fig.~\ref{fig:lc11072}, we have plotted the color curves of the region. These are derived by measuring the number of pixels in each map, which, in the RGB coordinate system, are found within a Euclidean distance of 60 from the coordinates given in Table~\ref{table:1}. For the following discussion it should be kept in mind that the color curves represent, essentially, not a flux content but the the number of pixels where combinations of the components are maximized according to Table~\ref{table:1}, and hence the area covered by those colors. We will not discuss the black color because its variation depends on the variable background noise of the magnetogram and it is only indirectly related to the evolution of the region, since it represents pixels where all three components are close to zero.  

Before discussing the color curves, we provide the evolution of the three components of the magnetic field as a reference (Fig.~\ref{fig:lc11072}a). They exhibit the typical evolutionary trends relevant to flux emergence and decay \citep[see e.g.,][and references therein]{2017Norton} with a brief increase, a short plateau, a more intense increase which, for this particular case led to a local maximum at around $t=1.5$\,days and a global maximum at $t = 2.2$\,days. The magnetic flux of the vertical component is higher while the two horizontal ones exhibit almost the same magnetic flux. The peak magnetic flux is followed by a slow decline, evident in all three components. The increase observed after central meridian distance 40\degr\ is a systematic effect due to the asymmetric noise pattern of HMI \citep{bobra14,2014SoPh..289.3483H}.
    
The color curves in Fig.~\ref{fig:lc11072}b and c contain more intricate information regarding the evolution of the emerging region. The number of white pixels (Fig.~\ref{fig:lc11072}b, black curve) generally increases but with deviations, reaching a maximum value at $t=3$\,days, right after all components reached their maxima. The red curve has overall higher values in comparison to the green and blue and it exhibits discrete peaks at the start of the observations as well as at $t=1.2$\,days and $t=2.2$\,days indicating emergence events. Thus, in the beginning most of the region was occupied by mainly horizontal magnetic field and the prevalence of red over blue is due to the orientation of the emerging bipole, which aligned very well with the $x$-axis of the SHARP cut-out (see also Fig.~\ref{fig:evo11072}a and b). The blue curve, on the other hand, exhibited the lowest values at all times, as well as the most subtle variation because pixels with purely $B_{y}$ magnetic field were few and far between. It balanced the red one after $t=2.5$\,days, when the symmetric spot formed and then it overtook the red one after $t=4$\,days. At that point, no more horizontal flux was injected and thus the prevalence of red had ceased. The green curve exhibited a continuous increase during the entire span of observations, with a distinct peak at around $t=0.8$\,days, at the onset of the main emergence phase (see also Fig.~\ref{fig:evo11072}a) and another one at $t=1.8$\,days during the local maximum of the vertical magnetic flux. It continued to increase rather monotonously after $t=3$\,days until the region reached the limb, where it was also affected by projection effects. During its late evolutionary stages, the region consisted of a plage and a well-formed spot. The green curve dominated, and it was more than twice as high as the red and blue curves combined because most of the active region area contained pixel closer to the vertical direction whereas purely $B_{x}$ and $B_{y}$ fields occupied far fewer pixels. During the decay phase, this development also had an impact on the cyan, magenta, and yellow curves (Fig.~\ref{fig:lc11072}d).
   
Since in the RGB color system the cyan, magenta and yellow are secondary colors, the corresponding color curves represent the number of pixels where two out of the three components reach maximum values (Table~\ref{table:1}) and the third one is minimal. All three curves (Fig.~\ref{fig:lc11072}d) exhibit peaks in the beginning of the observations. then yellow and cyan increase continuously, with the former peaking at around $t=2.2$\,days and the latter saturating already after the first day. The yellow curve declines after the peak, whereas the cyan one is almost constant throughout the observations. The magenta curve exhibited distinct peaks, which coincided with peaks in red and green (since magenta represents pixels with both $B_{x}$ and $B_{y}$ maxima) and then acquired again very low values. 

\begin{figure*}[htp!]
\centering
\includegraphics[width=\hsize]{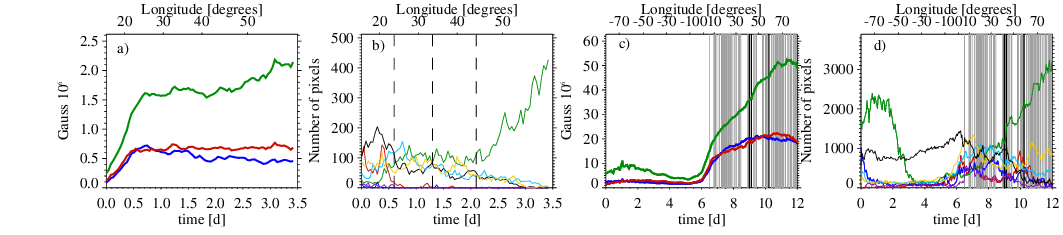}
\caption{ (a -- b) Magnetic flux evolution and color curves for active region NOAA\,12118. The vertical dashed lines mark the transitions between different evolutionary stages (see text). (c -- d) Magnetic flux evolution and color curves for active region NOAA\,12673. Vertical lines indicate flares and thick vertical lines indicate X-class flares. The time is in decimal days since the beginning of their observation, that is, at 17~July~2014 16:34\,UT for NOAA\,12118 and at 28~August~2017 08:58\,UT for NOAA\,12673.}
\label{fig:emergence_decay}
\end{figure*}

In summary, the temporal variation of the color curves appears to reflect important stages on the evolution of active regions, such as emergence, transition to a more relaxed or potential state, and decay. To further support the diagnostic value of COCOMAGs for flux emergence, we also present two well-studied examples of active regions, NOAA\,12118 and NOAA\,12673 (Fig.~\ref{fig:emergence_decay}). The former was studied by \citet{2016A&A...596A...3V} from its emergence to its decay, whereas the latter produced the strongest flare of Solar Cycle 24 and its evolution has been extensively studied \citep[see][for a comprehensive review of the literature]{Pietrow23}.

In Fig.~\ref{fig:emergence_decay}a and b we present the temporal evolution of $B_{x}$, $B_{y}$, and $B_{z}$, as well as the six color curves, deduced from the COCOMAGs, for active region NOAA\,12118. They present similar evolutionary characteristics as the ones described earlier for NOAA\,11072, although NOAA\,12118 was much weaker. \citet{2016A&A...596A...3V} distinguished four phases in the lifetime of this active region, that is the emergence phase exhibiting a fast expansion (``a''), followed by a slowing down of the expansion (``b'') after $t=0.6$\,days, its halt starting at $t=1.3$\,days (``c'')  and the onset of decay at $t = 2.1$\,days (``d'') . The corresponding times were derived using 17~July~2014 at 16:34\,UT as a reference. Comparing these times with the temporal evolution of the colorcurves in Fig.~\ref{fig:emergence_decay}b, we note the following: (a) during the fast emergence all curves increased with a dominance of red and white, (b) The bipole expansion slowed down when green, cyan and yellow overtook red and white (t$\sim$0.5\,days), and (c) was slowed when green marginally overtook yellow and cyan curves ($t\sim1.3$\,days). (d) Finally the onset of decay coincided with green dominating and initiating a fast increase ($t\sim2.1$\,days), while all the other color curves declined. In essence, as the active region emerged and decayed the area occupied by magnetic field of different orientation changed accordingly. Horizontal magnetic flux was injected to the photosphere during emergence, which accumulated to form pores, exhibited some twist and gradually relaxed to a more vertical state during decay.  

Active region NOAA 12673 is another interesting example, substantiating the previous description. The region rotated into view having the appearance of an already evolved region with a well-formed sunspot and trailing plage. In this case, the dominance of green and white observed in (Fig.~\ref{fig:emergence_decay}d) is in agreement with the previous description. When the region was closer to the disc center, the white pixels dominated, since the region consisted of a well formed symmetric spot. With the intense emergence events of 2~September 2017 ($t\sim5$\,days) and the subsequent interactions between polarities \citep{2018Verma}, the area occupied by strong horizontal magnetic field increased substantially. During the same time, red, and more prominently yellow and cyan, exhibited a strong increase, as a result of the injection of new horizontal magnetic flux and the development of strong shear (see also top panel of Fig.~\ref{fig:examples}). Four days later, decay set on, while the region rotated out of view. This period of intense emergence and interaction was also accompanied by heavy and repeated flaring, culminating to two X-class flares on 9~September 2017. The implied relationship between the color curves, the magnetic complexity, and flaring will be discussed in the following sections.

These example demonstrate that the COCOMAG plots allow to visualize the typical structure of the magnetic fields during the different evolutionary stages of an active region. Initially, during the formation of the region, horizontal components of the magnetic field are dominant. This is in agreement with the simulations of \citet{Rempel2014}, which show that during the early stages of active region formation, horizontal fields rise toward the surface and emerge through the photosphere. Subsequently, these horizontal fields gradually evolve into a mixture of weak magnetic fields and more vertical fields \citep[see][and references therein]{2015LRSP...12....1V}. This stage is also well captured in the COCOMAG plots and subsequent color curves, where the dispersed vertical magnetic field dominate the decaying regions.

\subsection{Color curves and magnetic complexity}
\label{section:colorcurves_sharp}

\begin{figure*}[htp!]
\centering
\includegraphics[width=18cm]{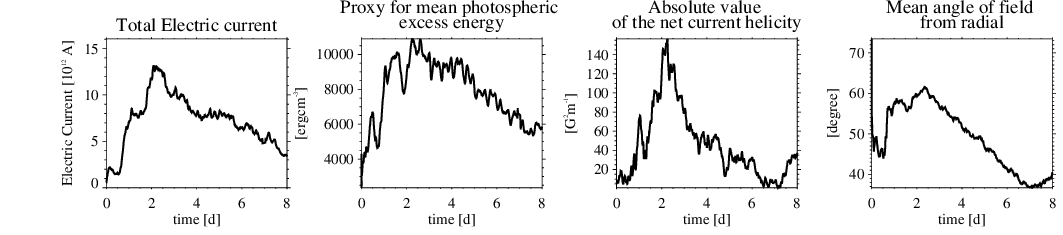}
\caption{ Temporal evolution of four selected SHARP parameters for the active region NOAA\,11072. From left to right: The total vertical electric current (TOTUSJZ), the proxy for mean photospheric excess energy (MEANPOT), the absolute value of the net current helicity (ABSNJZH), and the mean angle of the magnetic field from the radial (MEANGAM).}
\label{fig:sharp11072}
\end{figure*}

In Fig.~\ref{fig:sharp11072} we plot four selected magnetic properties, namely the integral of the total unsigned vertical electric current density, a proxy for the mean photospheric excess energy, the absolute value of the net current helicity and the mean angle of the magnetic field with respect to the radial direction, for active region NOAA\,11072. These four parameters, which are provided with the SHARP data product, quantify the non-potentiality of active regions and were chosen due to their relevance to fundamental physical quantities (electric current, magnetic energy, current helicity) and the ``geometry'' of the magnetic field (mean angle of the magnetic field). Their calculation formulas are given in \citet{bobra14} and they all involve combinations of the magnetic field components. 

All parameters exhibit an increasing trend up to a maximum value until $t=2.2$\,days, following the evolution of emerging flux and the trend indicated by the total unsigned magnetic flux. Notwithstanding this similarity, the four parameters exhibit different characteristics, which depend on their association with different aspects of complexity, which is a typical feature of these parameters \citep[see, e.g.][]{2018Kontogiannis}. Thus, the integrated vertical electric current and the current helicity tend to exhibit narrower peaks, due to their association with the $B_{x}$ and $B_{y}$, while the magnetic energy excess builds up rather continuously, although also exhibiting peaks. For all three parameters, these peaks are co-temporal with the peaks seen in the red and magenta color curves (Fig.~\ref{fig:lc11072}c,d). These peaks are more subtle for the mean angle of the field from the radial direction, which still resembles the yellow color curve (Fig.~\ref{fig:lc11072}d). This is reasonable since the yellow curve represents the area with pixels where $B_{x}$ and $B_{z}$ are maximum, for which the vector of the magnetic field can deviate significantly from the vertical. 

\begin{figure*}[htp!]
\centering
\includegraphics[width=18cm]{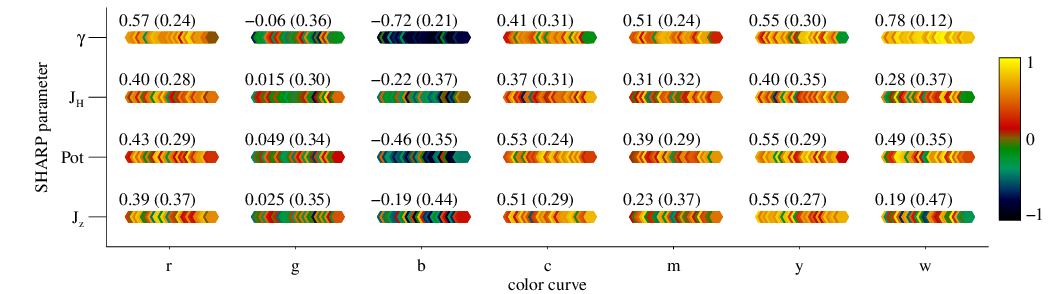}
\caption{Correlations between the six color curves and the four parameters of Fig.~\ref{fig:sharp11072}, calculated for each active region. Each hexagon represents the individual correlation coefficient between parameter and color curve, coded according to the color bar. Printed on top of each set of correlation are the average and standard deviation for the active region sample.}
\label{fig:corrs}
\end{figure*}

Based on these observations we examine the correlations between the color curves and the four SHARP parameters of Fig.~\ref{fig:sharp11072} for each of the 33 active regions (see Table~\ref{table:2}). The correlations calculated for each region are presented in Fig.~\ref{fig:corrs}. The correlation coefficients between non-potentiality parameters and color curves differ, despite some systematic behavior. On average, we find a weak to moderate correlation between the red, cyan, magenta and yellow curves with the four parameters and in many individual cases this correlation is strong. The green and particularly the blue component are weakly correlated/anticorrelated with the parameters. The same geometrical reason invoked earlier on the orientation of the emerging bipole applies as well here and can also explain the strong anti-correlation with the $\gamma$-angle. The latter is, on average, strongly correlated with the white color curve, because it represents pixels where all magnetic field components are equally strong and hence, the magnetic field is inclined.

\subsection{Color curves and flaring activity}

Although active region NOAA\,11072 was a rather quiet region at the beginning of Solar Cycle 24, when it reached maximum complexity it produced some weak flaring activity. Out of the four B-class flares it produced in total, two of them took place during the third day, as indicated by the vertical dashed lines in Fig.~\ref{fig:lc11072}. Panels (c) and (d) of the same figure show that the two flares coincided with or followed peak values in the red, yellow, and magenta color curves, all reflecting changes in the horizontal component, in association with strong vertical magnetic field. Similarly, active region NOAA\,12673 (Fig.~\ref{fig:emergence_decay}d) exhibited its strong flaring activity after $t=6$\,days, with the onset of fast flux emergence. As discussed earlier, all color curves increased, exhibiting multiple peaks after that time. In the previous sections we established the response of emergence and also of the twisting/shearing of the magnetic field that ensued due to opposite polarity motions on behavior of the color curves, as well as a moderate correlation between some of the color curves and common non-potentiality parameters. Motivated by this observations, here we examine the potential use of the color curves as predictors of flaring activity.

\begin{figure*}[htp!]
\centering
\includegraphics[width=\hsize]{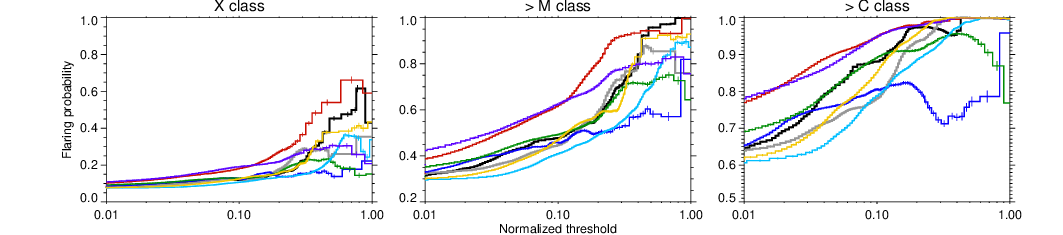}
\caption{Bayesian flaring probabilities and accompanying errors calculated for the 33 active regions in Table~\ref{table:2} as a function of normalized thresholds. Each color denotes probabilities calculated for the corresponding color curves (the gray color denotes probabilities for the ``w'' color curve).}
\label{fig:flare_probs}
\end{figure*}

We calculated the COCOMAGs and derived the color curves for the active regions of Table~\ref{table:2}. For each value of the color curve we measured the number of flares that took place within a window equal to 24\,h and belonged to the following categories, based on their magnitudes: higher than or equal to X1.0, higher than or equal to M1.0, and higher than or equal to C1.0. We then calculated the Bayesian flaring probabilities using the formulas presented in \citet{2018Kontogiannis}, for a set of thresholds chosen appropriately so each parameter bin contained the same number of values. The flaring probabilities were compared with the ones calculated for the time series of the total unsigned magnetic flux, which represents the most rudimentary complexity parameter. All thresholds are normalized to their maximum value to facilitate comparison between the different parameters and are presented in Fig.~\ref{fig:flare_probs}.

Three groups of color curves can be distinguished. Blue and cyan are associated with the lowest flaring probabilities. White, green and yellow exhibit higher flaring probabilities and are largely comparable with the total unsigned magnetic flux. Finally, red and magenta exhibit the highest probabilities, surpassing in most cases the total unsigned magnetic flux. Therefore, the area occupied by strong, purely horizontal magnetic field as quantified by the red and magenta color curves derived by the COCOMAGs could be used as a predictor of imminent flaring activity. Through their specific association to $B_{x}$ (red) and $B_{x}-B_{y}$ (magenta), these color curves are particularly sensitive to the emergence of new flux and to the development of twist and shear, which trigger intense activity.

We conclude that the color curves derived from COCOMAGs, quantify the part of the region occupied with magnetic field components of prescribed strength, and combine area, magnetic field strength and orientation information. This information is related to evolutionary aspects of the emerging magnetic flux and to the amount of the developed non-potentiality. These empirical facts, combined with their association with imminent flaring activity and their relatively simple derivation render COCOMAGs promising for probing the evolution of active regions and advancing the prediction of their erupting activity. 

\section{Conclusions and discussion}

In this study, we utilized HMI vector magnetograms to study magnetic flux emergence and active region evolution through a method that transforms the magnetic field vector into a vector in the RGB color space. The produced COCOMAGs, essentially color maps encapsulating all three Cartesian components of the magnetic field, represent valid magnetic field information in a concise way. The Cartesian components represent flux density and therefore are suitable for such a treatment, in contrast with spherical components, where one would have to combine both magnetic flux density and angles. In the context of our methodology, the RGB primary colors red, green, and blue represent regions where one of the three components, $B_{x}$, $B_{z}$, and $B_{y}$ prevail, while the secondary colors cyan, magenta, and yellow represent regions where two out of three components are significant. Measuring the area occupied by each color, we created the corresponding color curves, which can be used along with the COCOMAGs to probe active region evolution.  

A comparison with known, well-studied examples showed that the COCOMAGs and the derived color curves can be used as diagnostics of different evolutionary stages of active regions. The emergence of $\Omega$-loops is evident as a considerable area covered by horizontal magnetic field in between separating patches of rather vertical magnetic field. During the first days of emergence most of the region area is covered with moderately to highly inclined magnetic field, also developing some degree of twist, as the two polarities separate and build up. Higher degree of twist is indicated by more asymmetric colored patterns over sunspots as well as progression of colors in between, whereas shear is betrayed by thin strips of red and magenta colors and abrupt color transitions in regions of interactions. The initiation of the decay process can be determined as the time when most of the area of the region is covered by rather vertical magnetic field, indicated by prevailing green pixels. 

Motivated by this diagnostic capability of the color curves, we compared them with typical non-potentiality parameteres, such as the vertical electric current density, the free energy, the electric current helicity and the shear angle. A correlation analysis revealed a generally moderate correlation between the non-potentiality parameters and some of the color curves, particularly the red, magenta, cyan, yellow and white. This correlation indicates that the COCOMAG methodology offers an alternative to characterise active region complexity, since the color curve essentially represent a combination of geometric information of the photospheric magnetic field and area coverage. 

We examined the potential of the color curves as predictors of imminent flaring. The red and magenta color curves, which represent the area occupied by strong, purely horizontal magnetic field, exhibit the highest flaring probabilities, surpassing the flaring probabilities of the total unsigned magnetic flux. The yellow and white ones, which correspond to the area occupied by strongly inclined magnetic field also compete well with the total magnetic flux, while green, blue and cyan are associated with the lowest flaring probabilities. This is a first indication that there are prospects of adapting and including the COCOMAG methodology in active region complexity characterization and eruption prediction schemes using conventional statistics or machine learning (ML). 

Indeed, the abundance of solar observational data makes it an excellent candidate for ML processing. One of the most popular tasks is space weather prediction. One can distinguish between two types of forecasting schemes, i.e. image-based and parameter (feature) based. The latter are more common \citep[see][]{GEORGOULIS2024}, mostly for historical reasons and energy efficiency. Currently, however, image-based models are also gaining popularity for forecasting extreme events, such as vision transformers
\cite[e.g.,][]{2020arXiv201011929D,2024SoPh..299...33G}, deep convolutional neural networks \citep{2022PhRvR...4b3028J,9671322}, and autoencoders \citep{2022FrASS...937863W}. Due to their high information content, especially with respect to morphology, the COCOMAGs are a suitable candidate for some of the aforementioned algorithms. Furthermore, the by-product of COCOMAGs, i.e., color curves, could be used in conventional feature-based prediction schemes, such as Support Vector Machines (SVMs) and Long Short-Term Memory (LSTM) neural networks. As mentioned above, these show promising flare prediction potential, so a logical next step will be to implement them in such flare forecasting models along with existing empirical parameters.

Additionally, with the increasing abundance of multi-wavelength, multi-layer spectropolarimetric observations we expect that more height-resolved magnetic field products will become more available \citep{2019A&A...632A.112Y,2021A&A...647A.190B}. The COCOMAG methodology can be adapted to scrutinize that type of data, as well as to treat data cubes resulting from modelling or simulations \citep{2024ApJ...963L..21J}, bringing out subtle differences in the magnetic field per height and/or per component. Although the method relies on colors, it is a first step towards alternative representations of information regarding magnetic field. These representations can be adapted to user needs by choice of appropriate color tables, facilitate the transformation to various color systems, and further lead to products based on tactile perception and sonification.

\begin{acknowledgements}
     We would like to thank the anonymous referee for providing comments which contributed to the clarity of the manuscript. I. Kontogiannis is supported by the \emph{Deut\-sche For\-schungs\-ge\-mein\-schaft (DFG)\/} project number KO 6283/2-1. A.G.M. Pietrow and M.K. Druett are supported by the European Research Council (ERC) under the European Union’s Horizon 2020 research and innovationprogramme (grant agreement No. 833251 PROMINENT ERC-ADG 2018). E. Dineva thanks the Belgian Federal Science Policy Office (BELSPO) for the provision of financial support in the framework of the Brain-be program under contract B2/202/P1/DELPH and the Belgian Defence – Royal Higher Institute for Defence (RHID) through contract 22DEFRA006.
\end{acknowledgements}

%
%

\bibliography{references}{}
\bibliographystyle{aa}

\end{document}